\documentclass{Interspeech}

\interspeechcameraready 

\usepackage{multirow}
\usepackage{subcaption}
\usepackage{booktabs} 
\usepackage{caption}  
\usepackage{array}

\title{FUSE: Universal Speech Enhancement using Multi‐Stage Fusion of Sparse Compression and Token Generation Models for the URGENT 2025 Challenge}

\author[affiliation={1}]{Nabarun}{Goswami}
\author[affiliation={1,2}]{Tatsuya}{Harada}

\affiliation{}{The University of Tokyo}{Japan}
\affiliation{}{RIKEN}{Japan}
\email{\{nabarungoswami,harada\}@mi.t.u-tokyo.ac.jp}

\usepackage{comment}

\begin{document}

\maketitle

\begin{abstract}
We propose a multi-stage framework for universal speech enhancement, designed for the Interspeech 2025 URGENT Challenge. Our system first employs a Sparse Compression Network to robustly separate sources and extract an initial clean speech estimate from noisy inputs. This is followed by an efficient generative model that refines speech quality by leveraging self-supervised features and optimizing a masked language modeling objective on acoustic tokens derived from a neural audio codec. In the final stage, a fusion network integrates the outputs of the first two stages with the original noisy signal, achieving a balanced improvement in both signal fidelity and perceptual quality. Additionally, a shift trick that aggregates multiple time-shifted predictions, along with output blending, further boosts performance. Experimental results on challenging multilingual datasets with variable sampling rates and diverse distortion types validate the effectiveness of our approach.
\end{abstract}

\section{Introduction}

Speech enhancement is a fundamental task in audio processing with applications ranging from telecommunications to assistive technologies \cite{loizou2013speechenhancement}. The increasing demand for systems that perform robustly in real-world conditions has driven research into models that generalize across multiple distortion types, languages, and sampling rates. In this context, the Interspeech 2025 URGENT Challenge\cite{URGENT2025}\footnote{\url{https://urgent-challenge.github.io/urgent2025/}} provides an ideal platform to evaluate universal, robust, and generalizable solutions under adverse conditions.

Traditional speech enhancement techniques have evolved from classical methods such as spectral subtraction and Wiener filtering to deep learning models \cite{defossez2020music, tong2024scnet, wang2023tf}, which have significantly advanced source separation and enhancement capabilities. Recent advances in generative enhancement methods have further broadened the field. Approaches based on diffusion models \cite{richter2023speech, lemercier2023storm}, generative adversarial networks (GANs) \cite{su2021hifi, babaev2025finally}, and token-based generative models \cite{yang24h_interspeech} have shown promise in improving perceptual quality by modeling the distribution of clean speech. Token-based methods exploit discrete acoustic representations from neural audio codecs \cite{kumar2024high, defossez2023high} to reconstruct missing information. Moreover, token-based generative modeling has achieved high fidelity in speech synthesis tasks, with methods like SoundStorm \cite{borsos2023soundstorm} and Injection-Conformer\cite{goswami2025edmtts} pushing the limits of realistic speech generation. Various self-supervised learning (SSL) models \cite{chen2024xeus, chen2022wavlm} have also been proposed recently which have enabled robust speech feature extraction. However, typical discriminative source separation based methods struggle with perceptual quality while generative methods suffer from drops in signal level intrusive metrics. TokenSplit \cite{erdogan23_interspeech} uses a 2-stage hybrid approach but relies on text transcripts, which may not always be available.

Motivated by these developments and challenges, we explore a hybrid approach that integrates discriminative and generative paradigms. Our multi-stage framework begins with a Sparse Compression Network (SCNET) \cite{tong2024scnet} augmented with dual-path Conformer layers \cite{gulati20conformer} to perform robust source separation and obtain an initial clean speech estimate. This is followed by an efficient, generative model—the Injection-Conformer \cite{goswami2025edmtts}—adapted to speech enhancement by leveraging SSL features and a masked language modeling objective on acoustic tokens from a neural audio codec. Finally, a fusion network aggregates the outputs of the previous stages with the original noisy input, striking a balance between intrusive and perceptual metrics. Additional techniques, such as a shift trick for time equivariance \cite{defossez2020music} and network blending \cite{uhlich2017blending}, further enhance the overall system robustness.

\noindent Our key contributions are:
\begin{itemize}
    \item A novel multi-stage fusion framework that integrates discriminative and generative models to balance signal fidelity and perceptual quality and achieved third rank in the URGENT 2025 challenge.
    \item Adaptation of a generative model to speech enhancement by leveraging SSL features and a masked language modeling objective on neural audio codec tokens.
    \item A fusion network that combines outputs from multiple stages.
\end{itemize}
The remainder of the paper is organized as follows. Section \ref{sec:proposed} details the proposed multi-stage framework, Section \ref{sec:exp} presents experimental evaluations, and Section \ref{sec:conclusion} concludes with a discussion of results, limitations, and future directions.

\section{Method} \label{sec:proposed}

\begin{figure*}[t]
    \centering
    \begin{subfigure}[b]{0.68\textwidth}
        \centering
        \includegraphics[width=\textwidth]{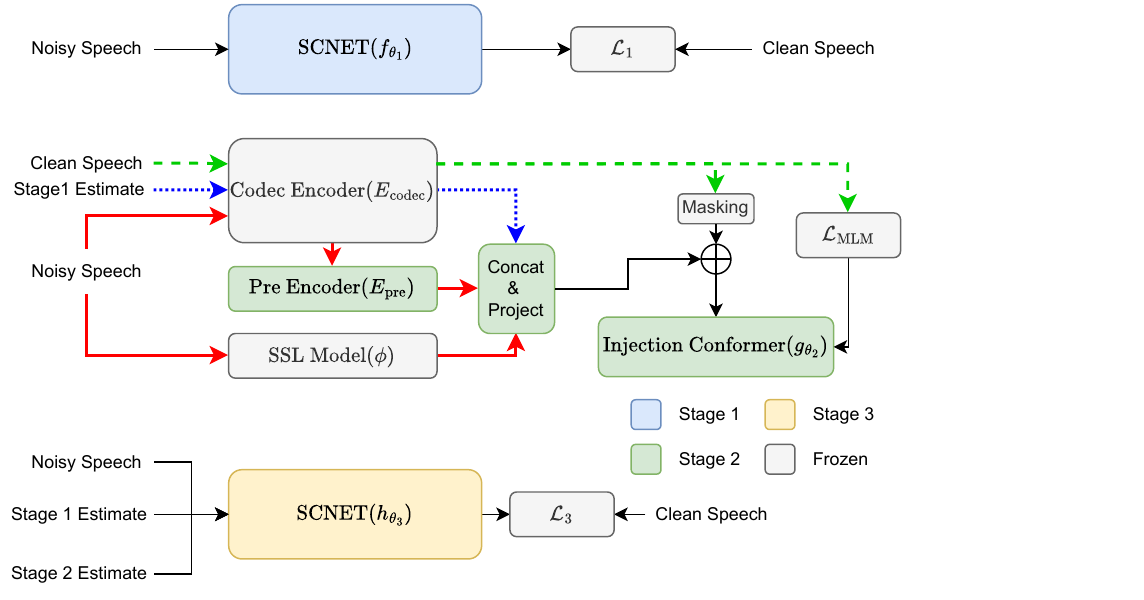}
        \caption{Training pipeline}
        \label{fig:sub1}
    \end{subfigure}
    \hfill
    \begin{subfigure}[b]{0.3\textwidth}
        \centering
        \includegraphics[width=\textwidth]{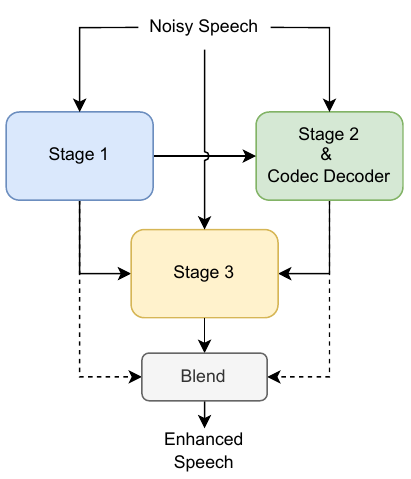}
        \caption{Inference pipeline}
        \label{fig:sub2}
    \end{subfigure}
    \caption{Model architectures and training and inference pipeline illustration.}
    \label{fig:main}
\end{figure*}
In this section, we detail our multi-stage speech enhancement framework that progressively refines noisy speech through three key stages: a Sparse Compression Network, a Token Sampling Generative Model, and a Fusion Network. 
The following subsections describe each component of our method in detail and Fig. \ref{fig:main} provides a visual overview of the training and inference pipelines of our method.

\subsection{Stage 1: Sparse compression network}
Let \( x \in \mathbb{R}^{T} \) denote the noisy speech signal, where \( T \) is the number of time samples. In this stage, we train a Sparse Compression Network (SCNET) \cite{tong2024scnet} to generate an initial enhanced estimate \( \hat{s}_1 \in \mathbb{R}^{T} \) from \( x \) via the transformation:
\begin{equation}
\hat{s}_1 = f_{\theta_1}(x),
\end{equation}
where \( f_{\theta_1}(\cdot) \) is parameterized by \( \theta_1 \).
SCNET partitions the input spectrogram into subbands and employs a sparsity-based encoder with adaptive compression to enhance separation performance while reducing computational cost.
It employs a dual-path BLSTM stack as the backbone. To support online usage and facilitate subsequent training stages, we replace the original BLSTM layers in SCNET with Conformer layers \cite{gulati20conformer} for improved parallelizability.

The training objective for Stage 1 combines a multi-scale mel-spectrogram loss \( \mathcal{L}_{\text{mel}} \) \cite{kumar2024high} with a Scale-Invariant Signal to Distortion Ratio (SI-SDR) loss \( \mathcal{L}_{\text{SI-SDR}} \) \cite{le2019sdr}:
\begin{equation}\label{eq:s1loss}
\mathcal{L}_{1} = \mathcal{L}_{\text{mel}} + \mathcal{L}_{\text{SI-SDR}}.
\end{equation}
This discriminative model is effective at extracting the speech signal from noisy mixtures; however, under challenging conditions—such as high noise, packet loss, or limited bandwidth—its ability to preserve high-quality speech may be compromised.

\subsection{Stage 2: Token sampling based generative model}
Stage 2 refines the initial estimate \( \hat{s}_1 \) by adapting a generative model built on a residual vector quantization-based audio codec framework. Although we initially considered SoundStorm \cite{borsos2023soundstorm}, its training complexity led us to adopt the more efficient Injection-Conformer model recently proposed in \cite{goswami2025edmtts}. The Injection-Conformer model efficiently captures the conditional dependence among different acoustic quantization levels at an architectural level by utilizing hierarchical injections, which allow teacher forcing for stable and faster convergence during training and efficient inference. This stage is designed to enhance the perceptual quality of the speech and recover missing information that may not be captured by the discriminative Stage 1. 

In this stage, the model is conditioned on a combination of features derived from both the noisy input and the Stage 1 output. Specifically, we extract self-supervised learning (SSL) features \( \phi(x) \) from the noisy input, obtain codec encoder features \( E_{\text{codec}}(\hat{s}_1) \) from the Stage 1 output, and further process the codec encoder features of the noisy input \( E_{\text{codec}}(x) \) with a trainable pre-encoder \( E_{\text{pre}}(\cdot) \), composed of a stack of conformer layers. These features are concatenated along feature dimension and projected to form the conditioning vector:
\begin{equation}
c = \operatorname{Proj}\Bigl([\phi(x);\, E_{\text{codec}}(\hat{s}_1);\, E_{\text{pre}}(E_{\text{codec}}(x))]\Bigr),
\end{equation}
where \( \operatorname{Proj}(\cdot) \) is a learnable projection layer.
During training, the conditioning vector \( c \) is added to the embeddings of the masked tokens of the clean speech’s codec tokens. The generative model \( g_{\theta_2}(\cdot) \) then predicts the codec tokens for these masked positions:
\begin{equation}
\hat{z} = g_{\theta_2}(c + z_{\text{clean}}^{\text{masked}}),
\end{equation}
where \( z_{\text{clean}}^{\text{masked}} \) represents the embeddings of the masked codec tokens derived from the clean speech. The model is trained using a masked language modeling (MLM) loss defined via cross entropy:
\begin{equation}
\mathcal{L}_{\text{MLM}} = - \sum_{i \in \mathcal{M}} \log P_{\theta_2}(z_i \mid \hat{z}_i),
\end{equation}
with \( \mathcal{M} \) denoting the indices of the masked tokens and \( z \) representing the ground-truth codec tokens of the clean speech. During Stage 2 training, the parameters \( \theta_1 \) remain frozen. During inference, the codec decoder \( D(\cdot) \) is employed to convert the predicted tokens into a waveform. While it improves speech quality, it can result in lower signal-level metrics, such as SDR.

\subsection{Stage 3: Fusion network}
Stage 3 integrates the outputs from Stages 1 and 2 to generate the final enhanced speech \( \hat{s}_3 \). The fusion network \( h_{\theta_3}(\cdot) \) adopts the same architecture as Stage 1, with its input layer modified to accept three signals: the original noisy speech \( x \), the discriminative output \( \hat{s}_1 \), and the generative refinement \( \hat{s}_2 \) (obtained via non-iterative greedy decoding). This fusion is motivated by the desire to combine the high signal-level accuracy of Stage 1 with the perceptual improvements from Stage 2, thereby achieving balanced performance across both signal-level and quality-based metrics. Specifically, the fusion process is defined as:
\begin{equation} \label{eq:fusion}
\hat{s}_3 = h_{\theta_3}\Big( [x; \hat{s}_1; \hat{s}_2] \Big),
\end{equation}
where \([\,\cdot\,; \cdot\,]\) denotes channel-wise concatenation of the input signals.
The overall training loss for Stage 3 is the sum of the primary reconstruction losses (Eq. \ref{eq:s1loss}) and additional task-specific losses. The primary losses include the multi-scale mel-spectrogram loss \( \mathcal{L}_{\text{mel}} \) and the SI-SDR loss \( \mathcal{L}_{\text{SI-SDR}} \). In addition, we incorporate the speaker encoder cosine similarity loss, the phoneme encoder feature matching loss \cite{xu2021simple}, and the UTMOS loss \cite{saeki22utmos} (defined as \(5\text{-}\operatorname{UTMOS}(\cdot)\)). The total loss is given by:
\begin{equation} \label{eq:loss_stage3}
\mathcal{L}_{3} = \mathcal{L}_{\text{mel}} + \mathcal{L}_{\text{SI-SDR}} + \mathcal{L}_{\text{spk}} + \mathcal{L}_{\text{phoneme}} + \mathcal{L}_{\text{UTMOS}}.
\end{equation}
During Stage 3 training, the parameters \( \theta_1 \) and \( \theta_2 \) remain frozen, ensuring that only the fusion network \( h_{\theta_3}(\cdot) \) is updated.

\subsection{Shift trick and network blending}
We further improve robustness by addressing time equivariance using a shift trick inspired by Demucs \cite{defossez2020music}. For any stage output \( \hat{s}_i \) (with \( i = 1,2,3 \)), we generate a set of predictions shifted by different time offsets. Specifically, for each time shift \( \tau \) in the set \( \mathcal{T} \), we compute:
\begin{equation} \label{eq:shift}
\hat{s}_i^{(\tau)}(t) = \hat{s}_i(t+\tau), \quad \forall \tau \in \mathcal{T},
\end{equation}
where \( \tau \) denotes the time offset. An inverse shift \( \text{Shift}^{-1} \) is then applied to realign each prediction with the original time axis, and the aggregated output for stage \( i \) is obtained by averaging:
\begin{equation} \label{eq:aggregate}
\bar{s}_i = \frac{1}{|\mathcal{T}|} \sum_{\tau \in \mathcal{T}} \text{Shift}^{-1}\bigl(\hat{s}_i^{(\tau)}\bigr).
\end{equation}
In addition, we incorporate network blending \cite{uhlich2017blending} to further enhance the final output. A subset \( \mathcal{I} \subseteq \{1,2,3\} \) of the stage outputs is blended, with the final enhanced speech computed as:
\begin{equation} \label{eq:blend}
\hat{s}_f = \frac{1}{|\mathcal{I}|} \sum_{i \in \mathcal{I}} \bar{s}_i.
\end{equation}

\section{Experiments} \label{sec:exp}
\begin{table*}[htbp]
    \small
    \renewcommand{\arraystretch}{0.8} 
    \centering
    \caption{Evaluation of the different stages on the non-blind test set.}
    \label{tab:nonblind}
    \begin{tabular}{l|c|c|c|c|c|c|c|c}
        \toprule
        Method                & Shifts    & UTMOS            & DNSMOS  & \shortstack{NISQA} & PESQ & LPS & SDR & \shortstack{SpkSim} \\
        \midrule
        Baseline              & -    & 2.11             & 2.94   & 2.89  & 2.43  & 0.79  & 11.29 & \underline{0.80}     \\
        \midrule
        Stage 1 (S1)          & -     & 2.18            & 2.96   & 2.95  & 2.47  & 0.80  & 12.62 & 0.79    \\
        Stage 1               & 10    & 2.20            & 2.97   & 3.10  & 2.49  & 0.80  & 12.81 & 0.79    \\
        \midrule
        Stage 2 (S2)          & -     & 2.38            & 3.07   & 3.21  & 2.34  & 0.80  & 9.03  & 0.79     \\
        Stage 2               & 10    & 2.39            & \underline{3.08}   & 3.38  & 2.53  & 0.81  & 11.21 & 0.78    \\ 
        \midrule
        Stage 3 (S3)          & -     & \textbf{2.64}            & \underline{3.08}   & 3.49  & 2.55  & \underline{0.82}  & 12.53 & \underline{0.80}     \\
        Stage 3               & 10    & \underline{2.55}         & \underline{3.08}   & \underline{3.59}  & 2.57  & \underline{0.82}  & \underline{12.72} & \underline{0.80}     \\
        \midrule
        Blend: S1 \& S2        & -   & 2.34            & 3.03   & 3.36  & \underline{2.65}  & \underline{0.82}  & 12.48 & \textbf{0.82}     \\
        Blend: S1 \& S2 \& S3  & -    & 2.45           & 3.06    & 3.47  & 2.64  & \textbf{0.83}  & \textbf{12.85} & \textbf{0.82}    \\ 
        Blend: S2 \& S3        & -     & 2.50           & \textbf{3.09}    & \textbf{3.60}  & \textbf{2.66}  & \textbf{0.83}  & 12.43 & \textbf{0.82}   \\ 
        \bottomrule
    \end{tabular}
\end{table*}

\begin{table*}[htbp]
    \small
    \renewcommand{\arraystretch}{0.8} 
    \setlength{\tabcolsep}{2pt}
    \centering
    \caption{Comparison of our method with the official baseline and the top 2 methods from the challenge on the blind test set. Superscript \textsuperscript{\dag} indicates hybrid discriminative and generative methods, while \textsuperscript{*} indicates discriminative methods. The numbers in bracket next to method name indicate the final challenge rank.}
    \label{tab:blind}
    \begin{tabular}{l |c| c| c| c| c| c| c| c| c| c| c| c| c| c}
        \toprule
        Method & DNSMOS & NISQA & UTMOS & POLQA & PESQ & ESTOI & SDR & MCD↓ & LSD↓ & SBS & LPS & SpkSim & CER (\%)↓ & MOS \\
        \midrule
        Baseline\footnotesize{(10)}\textsuperscript{*} & 2.85 & 2.77 & 1.92 & 2.99 & 2.24 & 0.76 & 10.24 & \underline{3.80} & \textbf{2.72} & 0.82 & 0.67 & 0.70 & 75.60 & 2.96 \\
        \midrule
        System A\footnotesize{(1)}\textsuperscript{*} & 2.88 & 3.22 & 2.09 & \textbf{3.40} & \textbf{2.64} & \textbf{0.82} & \textbf{12.66} & \textbf{3.67} & \underline{2.93} & \textbf{0.87} & \textbf{0.74} & \textbf{0.76} & \textbf{79.80} & 3.24 \\
        System B\footnotesize{(2)}\textsuperscript{\dag} & \underline{2.92} & \underline{3.24} & \underline{2.16} & \underline{3.17} & \underline{2.47} & 0.79 & 11.10 & 3.96 & 2.99 & \underline{0.84} & 0.70 & \underline{0.74} & 76.06 & \underline{3.32} \\
        Ours\footnotesize{(3)}\textsuperscript{\dag} & \textbf{2.94} & \textbf{3.25} & \textbf{2.19} & 3.16 & 2.45 & \underline{0.79} & \underline{11.25} & 4.79 & 3.66 & 0.83 & \underline{0.71} & 0.71 & \underline{77.09} & \textbf{3.44}\\
        \bottomrule
    \end{tabular}
\end{table*}

\subsection{Model architecture}
For Stage 1, we adopt the same configuration as the SCNET large\footnote{\url{https://github.com/starrytong/SCNet}}, with the BLSTM modules replaced by conformer layers with 8 attention heads, a convolution kernel size of 3, and a feedforward multiplier of 2. For stage 2, we use the XEUS\cite{chen2024xeus} SSL model\footnote{\url{https://huggingface.co/espnet/xeus}}\cite{chen2024xeus} and the DAC\cite{kumar2024high} 44Khz codec\footnote{\url{https://github.com/descriptinc/descript-audio-codec}}. The Injection-Conformer (reproduced based on \cite{goswami2025edmtts}) consists of 20 conformer layers, with 8 injections at [4, 6, 8, 10, 12, 14, 16, 18]. The pre-encoder is a stack of 6 conformer layers. All conformer layers in stage 2 feature a hidden size of 1024, 16 attention heads, a feedforward multiplier of 4, and a convolution kernel size of 5. 
The Stage 3, the architecture is identical to that of Stage 1. 
Stage 1 and Stage 3 are trained at 48 kHz, while Stage 2 is trained at 44 kHz, as the DAC-44kHz codec is used. For Stage 1 and 3, Short-Time Fourier Transform (STFT) is computed with a 4096 point FFT with 75\% overlap with a 4096 point \textit{hann} window.
We use publicly available checkpoints of models used for additional losses: speaker encoder\footnote{\url{https://huggingface.co/Jenthe/ECAPA2}} for $\mathcal{L}_{spk}$, and phoneme recognizer\footnote{\url{https://huggingface.co/facebook/wav2vec2-xlsr-53-espeak-cv-ft}} for $\mathcal{L}_{phoneme}$

\subsection{Training infrastructure and settings}
Our method is implemented in PyTorch and executed on 4 NVIDIA A100 GPUs. For each stage the training is conducted for 100,000 iterations with an overall batch size of 64 with 2 second segments.
We use the Adam optimizer\cite{kingma2015adam} with hyperparameters set as follows: \(\beta_1 = 0.8\), \(\beta_2 = 0.99\), and \(\epsilon = 1 \times 10^{-8}\). The maximum learning rate is set at \(2.5 \times 10^{-4}\) and is linearly warmed up for the first 4,000 steps. Following the warmup, the learning rate decays according to a cosine schedule over the remaining iterations, reaching a minimum value of \(1 \times 10^{-6}\).
To accelerate training and reduce memory usage, we adopt \textit{bfloat16} based mixed-precision. However, operations requiring high numerical precision—such as the STFT, inverse STFT, and loss calculations—are performed in \textit{float32}.
Across all stages, exponential moving average (EMA) of the model weights is applied with a decay factor of 0.9995. During inference, a sliding window approach is applied with 50\% overlap to ensure smooth transitions between segments.

\subsection{Dataset and augmentation}
We utilize the training data provided for the Track 1 of the URGENT 2025 challenge, consisting of 2.5K hours of speech (5 languages: English, German, French, Spanish, and Chinese) and 550 hours of noise along with 60k simulated room impulse responses sourced from several publicly available datasets. The speech data has variable sampling rate based on the particular data source and estimated bandwidth. During training we create noisy signals on the fly by sampling from the speech and noise datasets. We additionally apply the following data augmentations: \textit{bandwidth limitation, clipping, codec (mp3, ogg), packet loss and simulated wind noise}\cite{Mirabilii2022windnoise}, which are the 7 target distortion types of the challenge. For validation we utilize the official validation set of the challenge comprising of 1000 simulated samples. The challenge provided two test sets, a non-blind set (ground truth available for offline evaluation) consisting of 1000 simulated samples and a blind test set (ground truth not provided) of 900 samples comprising of 50\% real samples. Additionally, the blind test set had an unseen language (Japanese) and some unseen distortions, with 150 samples per language. For further details of the data please refer to \cite{URGENT2025}.

\subsection{Evaluation metrics}
We utilize the official evaluation metrics defined by the challenge \cite{URGENT2025}. It consists of a comprehensive set of metrics that capture both objective and perceptual speech enhancement performance. In particular, non-intrusive metrics (DNSMOS, NISQA, UTMOS) assess perceptual quality without requiring reference signals, intrusive metrics: POLQA\cite{beerends2013perceptual}, PESQ\cite{rix2001perceptual}, ESTOI\cite{jensen2016algorithm}, SDR\cite{le2019sdr}, MCD\cite{mcd}, and LSD\cite{lsd} quantify signal fidelity and intelligibility using clean references, downstream-task-independent metrics: SpeechBERTScore (SBS)\cite{saeki2024speechbertscore} and Levenshtein Phoneme Similarity(LPS)\cite{lps} and downstream-task-dependent measures: Speaker Encoder Cosine Similarty (SpkSim) and Character Error Rate (CER) evaluate the impact on related applications. Finally, subjective quality is measured via Mean Opinion Score (MOS).

\subsection{Performance analysis of multi-stage fusion}
We evaluate the different stages of our multi-stage fusion framework on the non-blind test set using a combination of intrusive and non-intrusive metrics, as summarized in Table~\ref{tab:nonblind}. A few key observations can be made:
the discriminative Stage 1 model already delivers strong performance, surpassing the official TF-GridNet\cite{wang2023tf} baseline in all the metrics. Stage 1 excels in terms of signal-level reconstruction (e.g., SDR), however, its non-intrusive quality metrics remain suboptimal. In contrast, the generative Stage 2 enhances perceptual quality metrics (UTMOS, DNSMOS, NISQA) by reconstructing missing information, though this comes at the expense of a reduced SDR, which is expected given that the model is not optimized with a signal-level reconstruction objective.
Stage 3, designed to fuse the strengths of both approaches, achieves a well-balanced performance across all metrics by combining the high SDR of Stage 1 with the perceptual improvements of Stage 2. Moreover, the application of the shift trick further enhances performance, particularly boosting the SDR by more than 2dB in Stage 2 with 10 shifts. Finally, network blending, especially the combination of Stage 2 and Stage 3 outputs, yields the best overall results across most metrics, with the exception of UTMOS where Stage 3 excels due to its dedicated $\mathcal{L}_{\text{UTMOS}}$.

\subsection{Evaluation on the blind test set}
Table~\ref{tab:blind} compares our proposed method with the official baseline and the top two challenge submissions on the blind test set. Our approach, which combines discriminative and generative techniques, demonstrates competitive performance across a diverse set of metrics.

Notably, our method achieves the highest scores in DNSMOS, NISQA, UTMOS, and subjective MOS among the top 3 submissions, highlighting its superior perceptual quality. Although the top-ranked discriminative method (System A) excels in signal-level metrics such as POLQA, PESQ, ESTOI, and SDR, our hybrid approach strikes an effective balance by enhancing perceptual quality with only a modest compromise in objective signal fidelity. Furthermore, when compared to the second-ranked hybrid system, our method exhibits better perceptual quality while maintaining comparable performance on other metrics. These observations reinforce our finding that discriminative methods tend to perform well on intrusive metrics, whereas generative methods excel in perceptual metrics, with hybrid approaches achieving an optimal balance.
Overall, the results confirm that integrating generative refinement with discriminative enhancement yields a robust system that significantly improves the overall listening experience, making our approach a strong contender in challenging real-world scenarios.

\section{Conclusion and Limitations} \label{sec:conclusion}
In this paper, we introduced a multi-stage fusion framework for universal speech enhancement, tailored for Track 1 of the Interspeech 2025 URGENT Challenge. Our approach integrates a sparse compression network with dual-path Conformer blocks, a token sampling generative model based on the Injection Conformer architecture, and a fusion network that aggregates their outputs. Techniques such as shift trick for time equivariance and network blending further enhance performance, resulting in a balanced improvement across both intrusive signal-level and perceptual quality metrics.
Despite these promising results, our method has limitations. The sequential training and inference paradigm, along with the extra shift operations, currently restricts real-time applicability. Future work will aim to address these challenges and further optimize the system.

\section{Acknowledgements}
This work was partially supported by JST AIP Acceleration Research JPMJCR20U3, Moonshot R\&D Grant Number JPMJPS2011, JSPS KAKENHI Grant Number JP19H01115, and JP20H05556 and Basic Research Grant (Super AI) of Institute for AI and Beyond of the University of Tokyo.

\bibliographystyle{IEEEtran}
\bibliography{mybib}

\end{document}